
\input harvmac.tex
\noblackbox

\def\at{{\tilde \alpha}}

\def\sq2{\sqrt{2}}
\def\s42{ 2^{-{1\over 4} } }

\def\p{\partial}

\def\exp{{\rm exp}}

\def\ket#1{|#1\rangle}
\def\bra#1{\langle#1|}

\def\ra{\rightarrow}

\def\a{\alpha}

\def\cg{\cos (2\pi V)}
\def\sg{\sin (2\pi V)}
\def\cg2{\cos (\pi V)}
\def\sg2{\sin (\pi V)}
\def\cb2{\cos (\delta/2)}
\def\sb2{\sin (\delta/2)}

\def\betum{\delta}
\font\cmss=cmss10 \font\cmsss=cmss10 at 7pt
\def\IZ{\relax\ifmmode\mathchoice
   {\hbox{\cmss Z\kern-.4em Z}}{\hbox{\cmss Z\kern-.4em Z}}
   {\lower.9pt\hbox{\cmsss Z\kern-.4em Z}}
   {\lower1.2pt\hbox{\cmsss Z\kern-.4em Z}}\else{\cmss Z\kern-.4em
Z}\fi}

\lref\CLNY{C.~Callan, C.~Lovelace, C.~Nappi and S.~Yost,
Nucl.~Phys. {\bf B293} (1987) 83.}
\lref\CKLM{ C. G. Callan and I. R. Klebanov, Phys. Rev. Lett. {\bf 72}
(1994) 1968; C. G. Callan, I. R. Klebanov, A. Ludwig and J. M. Maldacena,
Nucl. Phys. {\bf B422} (1994) 417.}
\lref\PT{J. Polchinski and L. Thorlacius, Phys. Rev. {\bf D50} (1994) 622.}
\lref\gz{S. Ghoshal and A. B. Zamolodchikov, Int. J. Mod. Phys. {\bf A9}
(1994) 3841.}
\lref\fsw{P. Fendley, H. Saleur and N. P. Warner,
Nucl. Phys. {\bf B430} (1994) 577. hep-th/9406125}
\lref\fls{P. Fendley, A. Ludwig and H. Saleur, cond-mat/9408068}
\lref\fendley{P. Fendley, F. Lesage and H. Saleur, hep-th/9409176}
\lref\afflud{I. Affleck and A. Ludwig, Phys. Rev. Lett {\bf 67} (1991) 161.}
\lref\affludtwo{I. Affleck and A. Ludwig, Nucl. Phys. {\bf B352} (1991) 841.}
\lref\hof {D. R. Hofstadter, Phys. Rev. {\bf B14} (1976) 2239.}
\lref\wannier{G. H. Wannier, Phys. Status Solidi  {\bf B88} (1978) 757.}
\lref\cgcdef{C. G. Callan and D. Freed, Nucl. Phys. {\bf B374} (1992) 543.}
\lref\dissqm{A. O. Caldeira and A. J. Leggett, Physica  {\bf 121A}(1983) 587;
Phys. Rev. Lett. {\bf 46} (1981) 211; Ann. of Phys. {\bf 149} (1983) 374.}
\lref\osdqm{C. G. Callan, L. Thorlacius, Nucl. Phys. {\bf B329} (1990) 117.}
\lref\cardyone{J. L. Cardy,  Nuc. Phys. {\bf B240} (1984) 514.}
\lref\clny{C. G. Callan, C. Lovelace, C. R. Nappi, and S. A. Yost,
Nucl. Phys. {\bf B293} (1987) 83; Nucl. Phys. {\bf B308} (1988) 221.}
\lref\CTrivi{C. G. Callan, L. Thorlacius, Nucl.  Phys. {\bf B319} (1989) 133.}
\lref\fisher{M. P. A. Fisher and W. Zwerger, Phys. Rev. {\bf B32} (1985) 6190.}
\lref\klebanov{I. Klebanov and L. Susskind, Phys. Lett. {\bf B200} (1988) 446.}
\lref\ghm{F. Guinea, V. Hakim and A. Muramatsu,
Phys. Rev. Lett.  {\bf 54} (1985) 263.}
\lref\tseyt{E. Fradkin and A. Tseytlin, Phys. Lett. {\bf 163B} (1985) 123.}
\lref\abou{A. Abouelsaood, C. G. Callan, C. R. Nappi and S. A. Yost,
Nucl. Phys. {\bf B280} (1987) 599.}
\lref\kondo{I. Affleck and A. Ludwig, Nucl.Phys. {\bf B360} (1991) 641.}
\lref\caldas{C. G. Callan and S. Das,  Phys. Rev. Lett {\bf 51} (1983) 1155.}
\lref\cardytwo{J. L. Cardy, Nucl. Phys. {\bf B324} (1989) 581. }
\lref\isqm{C. G. Callan, ``Dissipative Quantum Mechanics in Particle Physics",
Princeton preprint PUPT-1350, in ``Proceedings of the Fourth International
Conference on Quantum Mechanics in the Light of New Technology",
S. Kurihara ed., Japan Physical Society (1992);
 C.L. Kane and M.P.A. Fisher, Phys Rev. Lett.{\bf 68} (1992) 1220. }
\lref\gsw{ M. B. Green, J. H. Schwarz, and E. Witten, ``Superstring Theory'',
Cambridge University Press (1987).}
\lref\freed{D. Freed, Nucl. Phys. {\bf B409} (1993) 565.}
\lref\calfelfrd{C.Callan, A. Felce, D. Freed,
Nucl.Phys.{\bf B392} (1993) 551.}
\lref\fendetal{P.~Fendley, A.~Ludwig and H.~Saleur}
\lref\azbel{M. Ya Azbel', Zh. Eksp. Teor. Fiz. {\bf 46} (1964) 929;
D. R. Hofstadter, Phys. Rev. {\bf B14} (1976) 2239;
G. H. Wannier, Phys. Status Solidi {\bf B88 } (1978) 757  }


\Title{\vbox{\baselineskip12pt
\hbox{PUPT-1528}\hbox{hep-th/9503014}}}
{\vbox{\centerline{Magnetic Fields and Fractional Statistics}
\vskip 4pt\centerline{in Boundary Conformal Field Theory}}}
\centerline{C.~G.~Callan, I.~R.~Klebanov, J.~M.~Maldacena and A.~Yegulalp}
\centerline{\it Department of Physics, Princeton University}
\centerline{\it Princeton, NJ 08544}
\vskip .3in
\centerline{\bf Abstract}
We study conformal field theories describing two massless one-dimensional
fields interacting at a single spatial point. The interactions we include
are periodic functions of the bosonized fields separately plus a ``magnetic''
interaction that mixes the two fields. Such models arise in open string
theory and dissipative quantum mechanics and perhaps in edge state tunneling
in the fractional quantized Hall effect. The partition function for such
theories is a Coulomb gas with interchange phases arising from the magnetic
field. These ``fractional statistics'' have a profound effect on the phase
structure of the Coulomb gas. In this paper we present new exact and
approximate results for this type of generalized Coulomb gas.

\smallskip

\Date{3/95}

\newsec{Introduction}

Free field theories in $1+1$ dimensions with boundary interactions of
various types have been the subject of much recent work. Such systems
describe impurity scattering \affludtwo, tunneling between
quantum Hall edges \fls , dissipative quantum mechanics  \refs{\dissqm,\osdqm}
 and, in particle physics, open strings in background fields
\refs{\clny,\abou,\tseyt} and monopole catalysis \caldas. These theories
are often soluble since all the nontrivial dynamics are localized at a single
spatial point and amount to a dynamical boundary condition on otherwise
free fields. As usual, an important problem is that of finding the
conformally-invariant fixed points to which these theories flow in the infrared
(often the limit of most physical interest). Even for the simple case of a
single free scalar field, it turns out that the class of conformally invariant
boundary conditions is much larger than the familiar Dirichlet and Neumann
conditions \CKLM . In this paper we will present a new example, interesting
both for its possible practical applications and for the theoretical issues it
raises, in which the conformal field theory is exactly soluble.

We will be concerned with the theory of two massless scalar fields on a half
line, with boundary interactions generated by a periodic potential for each
scalar field separately and by a magnetic field term which mixes the two
fields. This system describes in a very direct way the {\it dissipative}
quantum mechanics (DQM) of an electron moving in two dimensions subject to a
magnetic field and a square lattice potential (the Wannier-Azbel-Hofstadter
model \azbel ). By tuning the period of the potentials and the strength of the
magnetic field one can reach $c=2$ conformal theories corresponding to phase
transitions between localized and delocalized long-time behavior of the
electron. From the point of view of string theory, the conformal fixed points
identify spacetime field configurations that solve the classical open
string field equations.

In this paper we will construct such $c=2$ boundary conformal field theories.
Our results are a logical extension of our previous work \CKLM\
on the $c=1$ problem of a single scalar field subject to a boundary periodic
potential (and no magnetic field, of course). There we found that if the
period of the potential is chosen such that the boundary interaction has
naive dimension one, the theory is conformal for {\it any} potential strength.
The conformal boundary state, which summarizes the quite non-trivial scattering
of asymptotic particles from the boundary interaction, turns out to be a
global $SU(2)_L$ rotation (with rotation angle set by the strength of the
potential) of the non-interacting boundary state. The partition functions and
S-matrix elements of the critical theory are accordingly simple
and explicitly calculable.

The $c=2$ theory, the subject of this paper, is more complex and more
interesting. First, the magnetic field affects the dimension of the potential
operators. The condition that this dimension be unity defines, to lowest
order in the potential strength, a circle in the plane of magnetic
field strength, $\beta$, and strength of dissipation, $\alpha$ (this parameter
can also be thought of as controlling the period of the potential).
The interesting complexity comes in when we ask what happens
as the potential strength is turned on. We find that at certain isolated
points along this circle (``magic'' values of the magnetic field
and dissipation strength), the
situation is much like the $c=1$ case: the dimension of the potential operator
is unaffected by the strength of the interaction and the boundary state can
be explicitly constructed out of $SU(2)_L$ actions on the non-interacting
boundary state. At generic values of $\alpha$ and $\beta$, however, there are
subleading corrections to the beta function which cause a renormalization
group flow of the potential strength. For
$\alpha,\beta$ near the critical circle
 the potential strength $V$ flows to a perturbatively reliable infrared
stable fixed point at small $V_c$. These fixed points should have quite
interesting physics which can be explored by perturbative methods.

The theoretical framework for all of this is the one-dimensional Coulomb gas.
When the partition function for theories of the type we are discussing is
expanded in the potential strength, each insertion of a periodic potential
behaves like a charge interacting with other charges via a logarithmic
potential. The $c=1$ theory describes a gas of $+$ and $-$ charges,
with all orderings of the charges contributing with equal weight.
The $c=2$ case corresponds to a Coulomb
gas with two flavors ($X$ and $Y$) of charge. The two flavors are
independent, except for a phase factor $e^{i q_X q_Y\phi}$
associated with the interchange of an $X$ and a $Y$ charge,
where $q_X$ and $q_Y$ are the signs of the charges, and $\phi$ depends
on $\alpha$ and $\beta$.  Many interesting
effects have to do with these ``fractional statistics''
of the Coulomb gas (the ``magic'' points correspond to phases which are
integral multiples of $2\pi$, so that the two flavors become independent).
This Coulomb gas with ``fractional statistics'' has received little attention
to
date, but, as we hope this paper will demonstrate, it is worth investigating
as it may underlie some systems of real practical interest.

\newsec{Setup and Renormalization Group Analysis}

We will be considering a system of two massless scalar fields in $1+1$
dimensions. The fields are free in the bulk but have
certain boundary interactions. Specifically, we take the action to
be $S=S_{bulk}+S_{pot}+S_{mag}$ where
\eqn\action{\eqalign{
S_{bulk}&= {\alpha\over 4\pi}\int_0^T dt\int_0^l d\sigma
        \left( (\partial_\mu X)^2 +(\partial_\mu Y)^2 \right)\ , \cr
S_{pot}&= V \int_0^T dt\left( \cos X(0,t)+ \cos Y(0,t)\right)\ , \cr
S_{mag}&= i{\beta\over 4\pi}\int_0^T dt
        \left( X \partial_t Y
-Y \partial_t X \right)_{\sigma=0}~.}
}
The parameters $\alpha$ and $\beta$ are defined as in \calfelfrd:
$\alpha$ determines the strength of dissipation in DQM (or, if we rescale
the fields, the period of the interaction potential); $\beta$ is related
to the magnetic field normal to the $X-Y$ plane via
$\beta = 2\pi B$. We have normalized the fields so that the period of the
boundary potential is $2\pi$. At this point
we have chosen to put the boundary interactions on the $\sigma=0$ boundary
and to leave the $\sigma=l$ boundary free, but, depending on the application,
we might wish to put interactions on the other boundary as well or even
move the second boundary to infinity in order to discuss scattering.

Comparing with the action of the $c=1$ theory \CKLM, we note that the truly
new feature of the $c=2$ system is $S_{mag}$. This term introduces a
boundary coupling between $X$ and $Y$ without destroying the conformal
invariance. Indeed, for $V=0$, the theory is free and obviously
conformally invariant for any $\alpha$ and $\beta$. Turning on the
potential in general spoils the conformal invariance, but we will
find that it is preserved on some surfaces in the $(\alpha, \beta, V)$
space. The theory turns out to be especially simple
at the so-called ``magic'' parameter values,
$$ (\alpha, \beta)= \left ({1\over n^2+1}, {n\over n^2+1}\right )
\ ,\qquad\qquad n\in \IZ $$
where it is conformally invariant for {\it any} potential stength $V$.
Since the magnetic flux per unit cell of the potential lattice
is $2\pi\beta$, the ``magic'' points do not correspond to trivial fluxes
(although they do correspond to trivial interchange phases of the Coulomb gas).
Thus, these physically non-trivial points possess hidden simplicity
which can be efficiently explored using boundary state techniques.

The theory defined by \action\ makes perfect sense for non-compact
fields $X$ and $Y$ (this is the physically interesting choice from the
DQM point of view). In other applications we may define $X$ and $Y$
on circles of radius $R$. The periodicity of the potential constrains
$R$ to be a positive integer. A further constraint on the parameters
arises from the single-valuedness of the path integral.
Under a physically unobservable shift $X\to X+2\pi R$,
$$\Delta S_{mag} = i\beta R \int_0^T dt~ \partial_t Y= 2\pi \beta R^2
$$
where we take $Y$ to wind once around the target space. For the theory to
be well-defined, we require $\Delta S_{mag} = 2\pi n$, which implies
$ \beta= n/R^2 $. This is just the usual flux quantization condition.
Since both $R$ and $n$ are integers, only rational values of $\beta$
are admissible in compactified theories.

We shall want to compute the functional integral for this system. As is
well-known \cardytwo , it can be regarded either
as an open string partition function $Z^{B_VN}=tr(e^{-T H_{open}})$
or as the amplitude for {\it free} closed string propagation between two
boundary states $Z^{B_VN} = \bra{B_V} e^{- l (L_0 +\tilde{L}_0)} \ket N$.
In these expressions, the superscripts denote the boundary conditions at the
ends of the open string: $\ket N$ is the Neumann boundary state and
$\ket{B_V}$ is the boundary state induced by the magnetic and potential
interactions. Since the magnetic
boundary action taken by itself is exactly soluble \abou, it is useful to
express $\ket{B_V}$ as the magnetic boundary state, $\ket{B_0}$, acted on by
the
potential term in the path integral. This gives a more explicit expression
for the partition function on the cylinder of circumference $T$ and length $l$:
\eqn\partfun{
Z^{B_VN} = \bra{B_0} e^{-V\int_0^T dt \left[\cos X(0,t)+\cos Y(0,t)\right]}
e^{-l (L_0 +\widetilde{L}_0 ) } {\ket N}~.
}
To calculate this object, we expand it in powers of $e^{\pm iX}$ and
$e^{\pm iY}$:
\eqn\Cg{\eqalign{
&Z^{B_V N}= Z^{B_0 N}
\sum_0^{\infty}
{(-V/2)^{n^X_+ +n^X_-+ n^Y_+ +n^Y_-}
        \over n^X_+! n^X_-! n^Y_+! n^Y_-!}
\bigl\langle
\prod_{j=1}^{n^X_+} \int_0^T dt_j e^{iX(0,t_j)}\times\cr
&\prod_{j=1}^{n^X_-} \int_0^T dt_j e^{-iX(0,t_j)}
\prod_{j=1}^{n^Y_+} \int_0^T dt_j e^{iY(0,t_j)}
\prod_{j=1}^{n^Y_-} \int_0^T dt_j e^{-iY(0,t_j)}
\bigr\rangle~~.}
}
The angle bracket denotes expectation value in the soluble theory with
magnetic field on one boundary. This assigns a certain Gaussian propagator
to the $X(0,t)$ and $Y(0,t)$ fields. The partition
function then has the structure of a one-dimensional two-component plasma
with the $e^{\pm iX}$ and $e^{\pm iY}$ operator insertions acting as two
distinct flavors of $\pm$ charges.

We are mainly interested in finding the points in $(\alpha,\beta,V)$ space
where these theories are conformal. The cylinder partition function \partfun\
is a rather sensitive observable, but a simpler test of conformal invariance is
provided by its $l\to \infty$ limit. In this limit, the cylinder becomes
conformally equivalent to a disk and the partition function reduces to
$$
Z_{\rm disk}= \bra{B_V}   0\rangle =
\bra{B_0} e^{-V\int_0^T dt \left[\cos X(0,t)+\cos Y(0,t)\right]}
{\ket 0}~,
$$
where $T$ is the circumference of the outer boundary of the disk (it
will serve as an infrared cutoff for our calculations).
Its perturbative expansion is still given by the Coulomb gas formula
\Cg\ and the Green functions needed to evaluate it can be shown to have the
following simple forms \calfelfrd:
$$\eqalign{
&\vev{X(t) X(0)}= \vev{Y(t) Y(0)}=-{\alpha\over \alpha^2+\beta^2}
\ln \left (4\sin^2 {\pi t\over T} \right ) \cr
&\vev{Y(t) X(0)}={i\pi\beta\over \alpha^2+\beta^2} (-1+ 2\ {\rm frac}(t/T))
}
$$
where $0\le {\rm frac}(x)<1$ is the fractional part of $x$.
The $X$-charges (and the $Y$-charges) interact
with themselves via the usual Coulombic logarithmic potential, but the magnetic
field induces a peculiar interaction phase between different types of charge.
Since ${\rm frac}(x)$ jumps by unity every time $x$ passes through an integer,
the phase of the amplitude jumps by ${2\pi\beta/( \alpha^2+\beta^2)}$
when the time ordering of an $X$ and a $Y$ charge is interchanged.
The smooth linear variation of the phase between such jumps actually cancels
out in neutral amplitudes (the only ones of concern to us). Thus, the entire
effect of the $XY$ interaction is summarized by weighting different orderings
of the $X$ and $Y$ charges with appropriate phases: fractional statistics
in one dimension.

It is now fairly straightforward to work out the first two terms in the
renormalization group flow equation for the potential $V$. The first
term, coming from normal ordering a single insertion of the potential, reads
$$
{d V\over d\tau} =\left (1-{\alpha\over \alpha^2+\beta^2}\right)V\ ,\qquad
 \tau=\log (T/\epsilon)
$$
where $T$ is the disk circumference and $\epsilon$ is an ultraviolet cut-off.
In other words, the effective dimension of the boundary potential is
$ h= \alpha/( \alpha^2+\beta^2)$ and, to leading order, the theory is
conformally invariant on the circle $h=1$. In the $c=1$ case the leading
order condition for conformal invariance turned out to be exact.
In order to study whether the same is true in the $c=2$ case,
we need to calculate corrections to the beta function. Techniques for such
calculations were outlined in \klebanov.

The next correction is of $O(V^3)$ and is extracted from the correction to
matrix element of the $e^{iX(0)}$ operator coming from
insertions of $e^{iY(t_1)}$ and $e^{-iY(t_2)}$ near $e^{iX(0)}$
(with subsequent integration over $t_i$). If this has a logarithmic divergence,
we must interpret this as a correction of order $V^3$ to the beta function.
Isolating the connected part of the correlation function, and
including the phases that account for the different orderings of  $e^{\pm iY}$
with respect to $e^{iX}$ we find, after mapping disk to
upper half-plane, the integral
$$
{V^3\over 16 \pi^2}\left [\exp \left({2\pi i\beta\over \alpha^2+
\beta^2} \right)
+\exp \left(-{2\pi i\beta\over \alpha^2+\beta^2}\right)-2\right ]
\int_0^\infty dt_1 \int_{-\infty}^0 dt_2~ (t_1-t_2)^{-2}
$$
which is easily seen to have a logarithmic divergence:
$$
\int_0^\infty dt_1 \int_{-\infty}^0 dt_2~ (t_1-t_2)^{-2}
\rightarrow\int_\epsilon^T dt_1 {1\over t_1}= \log (T/\epsilon)~.
$$
The resulting corrected beta function equation is
\eqn\corrbeta{
{dV\over d\tau}= V\left (1-{\alpha\over \alpha^2+\beta^2}\right ) -
{V^3\over 4\pi^2}
\sin^2 \left({\pi \beta\over \alpha^2+\beta^2}\right )+ \CO(V^5) \ .
}

Now that we have computed the flow equation \corrbeta, we can study the
fixed points. There is a class of ``trivial'' fixed points where the beta
function vanishes order by order in $V$. The first term vanishes everywhere
on the $h=1$ circle, while the second term vanishes only at the discrete points
where ${\beta\over \alpha^2+\beta^2}$ is integer as well. These are precisely
the ``magic'' points where the interchange phases between the $X$ and $Y$
charges become trivial. On a disk, then, the $X$ and $Y$ charges no longer
interact,
so that the partition function reduces to a product of two $c=1$ partition
functions with marginal boundary potentials. The exact conformal invariance
of such $c=1$ theories was demonstrated in \refs{\CKLM,\PT}.
This implies that at the ``magic'' points on the $h=1$ circle the beta function
indeed vanishes order by order in $V$.

While the ``magic'' points are trivial fixed points for any $V$, the rest
of the $h=1$ circle has no true fixed points except for the free theory with
$V=0$. Consider, however, values of $\alpha$ and $\beta$ slightly outside
the circle: \corrbeta\ shows that, for
$h={\alpha/( \alpha^2+\beta^2)}$ slightly less than 1, $V$ flows to an
infrared stable fixed point at
$$
V_c^2 = 4\pi^2 (1-h)/\sin^2 \left({\pi \beta\over \alpha^2+\beta^2}\right)\ .
$$
The parameters $\alpha$ and $\beta$ are not renormalized and may be
chosen so that $V_c$ is small and the perturbative argument is reliable.
A most interesting feature of these non-trivial fixed points is that the flow
starts with a perturbatively relevant operator of dimension
$h<1$, which destabilizes the trivial fixed point at $V=0$. The effect of our
interchange phases is to turn a naively massive theory into a conformal one!
Could there be an experimental realization for this?

The picture of the critical surface in $(\alpha,\beta,V)$ space which emerges
from this is quite interesting. It is easiest to describe in terms of
slices through the surface at constant $V$. For $V=0$ we find
the circle ${\alpha/( \alpha^2+\beta^2)}=1$. As we increase $V$, the
circle bulges outwards, while staying pinned at the ``magic'' points.
In next-to-leading approximation, the critical circle deforms to the curve
$$
\beta(\alpha)^2={1\over 4}-\left (\alpha-{1\over 2}\right )^2+
{V^2\over 4 \pi^2}\alpha\sin^2 \left (\pi\sqrt{1-\alpha\over\alpha}\right )
\ .$$
It is not obvious from these arguments what happens as $V$ gets large,
but some sort of perturbation theory in $1/V$ ought to be possible
since, in that limit, one is approaching the trivial Dirichlet boundary
condition.

There are two instructive consistency tests to be made at this point. First
in open string theory, it is legitimate to regard the beta function as the
equation of motion following from an action function given by the disk
partition function itself: $S_{eff}=\ln Z_{\rm disk}$. The $V$-independent
piece, the disk amplitude in the presence of a boundary magnetic field,
was calculated a long time ago \refs{\abou,\tseyt}\ (it is the Born-Infeld
action). If we introduce independent potential strengths for $\cos X$ and
$\cos Y$, $V_X$ and $V_Y$, then there is a quadratic term given by
$${V_X^2 + V_Y^2\over 4} \int_0^{2\pi} {dy\over 2\pi} [2\sin (y/2)]^{-2 h}
$$
where $h=\alpha/(\alpha^2+\beta^2)$ is the dimension of the
potential operators, and we have set the circumference $T$ of the disk
to unity. This integral can be calculated \fendley, giving
$$ {V_X^2 + V_Y^2\over 4} {\Gamma (1-2h)\over \Gamma^2 (1-h)}=
{V_X^2 + V_Y^2\over 8} [h-1 + \CO((h-1)^2)]
$$
where the right hand side is an approximation valid near
the $h=1$ circle. The quartic terms have been calculated
for other reasons in \calfelfrd\ and, when added to the lower-order terms give,
in the vicinity of the $h=1$ circle,
$$\ln Z_{\rm disk}=\half \ln [1+ (\beta/\alpha)^2]+
{V_X^2 + V_Y^2\over 8} \left[{\alpha\over \alpha^2+\beta^2}-1 \right]
+ {V_X^2 V_Y^2\over 32 \pi^2}
\sin^2 \left({\pi \beta\over \alpha^2+\beta^2}\right )~.
$$
Interpreting this as an effective action for $V_X$ and $V_Y$, varying with
respect to $V_X$ and setting $V_X=V_Y$, we recover
precisely the  beta function \corrbeta !

Second, recall that in theories with boundary interactions, $\ln Z_{\rm disk}$
is the analogue of the Zamolodchikov $c$-function and is believed to
decrease along the renormalization group flow \afflud. It is not hard to check
that this is the case for our example. In the UV theory with
$V_X= V_Y=0$, $\ln Z_{\rm disk}=\half \ln [1+ (\beta/\alpha)^2]$.
It is also true that
$${\partial \ln Z_{\rm disk}\over\partial V_X} < 0\ ,\qquad\qquad
{\partial \ln Z_{\rm disk}\over\partial V_Y} < 0\ ,
$$
for $0 < V_X, V_Y < V_c$, which implies that $\ln Z_{\rm disk}$
indeed decreases along the trajectories. At the IR fixed point
with $V_X= V_Y=V_c$ it reaches the value
$$\ln Z_{\rm disk}=\half \ln [1+ (\beta/\alpha)^2]-
\left[{\alpha\over \alpha^2+\beta^2}-1 \right]^2
{\pi^2\over 2
\sin^2 \left({\pi \beta\over \alpha^2+\beta^2}\right )}\ ,
$$
to the order we have retained.

\newsec{Exact Boundary States at ``Magic'' Magnetic Fields}

In the previous section we showed that at certain isolated points (where
$\alpha=\alpha^2+\beta^2$ and $\beta/\alpha$ is integer)
the theory is exactly
conformal order by order in an expansion in the potential strength. We
will now show how to reduce the calculation of the boundary state at
these points to an algebraic exercise. Once the boundary state is known,
any other quantity of physical interest can be computed.

We start with the familiar oscillator expansion of the pure magnetic field
boundary state \abou:
$$\eqalign{
\ket{B_0} = ~\sqrt{1+(\beta/\alpha)^2}
&\exp~\bigl\{ \sum_{n>0}{1\over n} ( \at^X_{-n} (\cos \betum  \a^X_{-n}+
\sin \betum \a^Y_{-n}) +\cr
& \at^Y_{-n} (\cos \betum  \a^Y_{-n} - \sin \betum  \a^X_{-n}))
\bigr\} ~\ket{0}~. }
$$
In this expression, the oscillators are defined by the usual expansion
of the boundary field
\eqn\xexpansion{
X(t,0) = \sqrt{ \alpha \over 2}
\sum_{n\not= 0} {1\over |n|} ( \a^X_n e^{i 2 \pi n t/T} + \at^X_n
e^{-i 2 \pi n t/T} )
}
and the reflection phase $\betum $ is defined by
\eqn\rotangle{ \sin \delta= {2\alpha\beta\over \alpha^2+\beta^2}\qquad
                {\rm or} \qquad \tan{\delta\over 2} = {\beta\over\alpha}~.}
Using this definition of $\delta$, the overall normalization factor of
${\sqrt{1+(\beta/\alpha)^2}}$ can be rewritten as $\sec(\delta/2)$
and will be so written in all further appearances of the boundary state.
The net effect of the magnetic field is a chiral $O(2)$ rotation of the
Neumann boundary state:
$$
\ket{B_0} = \sec(\delta/2)  e^{i \betum  {\cal R}_L } \ket N
$$
where
$$
{\cal R}_L=  (y^0_L p^X_L - x^0_L p^Y_L) +
\sum_{n>0} {i\over n} ( \a^Y_n\a^X_{-n} - \a^Y_{-n}\a^X_n)~.
$$
(We have included the zero mode parts for completeness, but they do not
contribute since the Neumann state on which it acts has zero momentum).
The interacting boundary state is obtained from the above by acting
on it with the boundary potential term:
$$
\ket{B_V} = \sec(\delta/2)
   e^{ -H_{pot}(X)  -H_{pot}(Y) } e^{i \betum  {\cal R}_L } \ket N .
$$
where  $ H_{pot}(X)  =  \int dt V \cos X(t,0) $.

We now show how to relate this to the explicitly known $c=1$ boundary state
of \CKLM: First, move the $O(2)$ rotation to the left to get
$$\eqalign{
  e^{i\betum  {\cal R}_L}   e^{-i\betum  {\cal R}_L}
        &e^{ -H_{pot}(X)  -H_{pot}(Y) } e^{i\betum  {\cal R}_L} \ket N =\cr
        &  e^{i\betum  {\cal R}_L}   e^{\{ e^{-i\betum  {\cal R}_L}
        ( -H_{pot}(X)  -H_{pot}(Y)) e^{i\betum  {\cal R}_L}\}}  \ket{N}~ .}
$$
Since the $O(2)$ generator acts only left-movers, we must break
fields up into their chiral components in order to evaluate the
rotated potential term:
$$
e^{-i\betum  {\cal R}_L} H_{pot} (X_L + X_R) e^{i\betum  {\cal R}_L} =
        H_{pot} (\cos\betum  X_L - \sin\betum  Y_L + X_R) ~.
$$
Since this operator acts directly on a Neumann boundary state, we
can use the Neumann boundary condition $ X_L \ra X_R $ to reduce this to
$$
H_{pot} (\cos\betum  X_L - \sin\betum  Y_L + X_L)  =
 H_{pot}( 2  \cos{\betum /2} (\cos{\betum /2} X_L - \sin{\betum /2} Y_L) )~.
$$
A similar expression gives the effect of the rotation on $H_{pot}(Y)$.
To simplify further still, we define rotated and rescaled coordinates
\eqn\prf{
 X'= \cos{\betum /2} ( \cos{\betum /2}~X -\sin{\betum /2}~Y )\ , \qquad
 Y'=\cos{\betum /2} ( \cos{\betum /2}~Y + \sin{\betum /2}~X ) \ .
}
in terms of which the boundary state becomes
\eqn\factorize{
\ket{B_V} = \sec(\delta/2) e^{i\betum  {\cal R}_L}
        e^{-H_{pot}(2 X_L') -H_{pot}(2 Y_L') }
                \ket{N^{X^\prime}}\ket{N^{Y^\prime}}~.}
We have written the Neumann boundary state as the product of an
$X^{\prime}$ and a $Y^{\prime}$ factor to emphasize that, apart from
the $O(2)$ rotation, \factorize\ has become the product of two $c=1$
boundary states. The rescaling of the fields changes the effective value
of $\alpha$ in the bulk action \action\ to
$\tilde\alpha= \alpha/ \cos^2 \betum /2 =(\alpha^2+\beta^2)/\alpha$.
On the critical circle $\tilde\alpha=1$ and $ H_{pot}(2X'_L)$ is
the integral of a
dimension-one chiral field.  As a result, the $X^{\prime}$ and $Y^{\prime}$
factors in \factorize\ become identical to the $c=1$ conformal boundary
states constructed in \CKLM. Because the $O(2)$ rotation commutes with
the Virasoro generators, \factorize\ is guaranteed to satisfy the
famous conformal boundary state condition $(L_n-\tilde L_{-n})\ket{B}=0$
\CLNY.

There are two subtleties to bear in mind: compactification and interchange
phases. In performing the $ X_L \ra X_R $ replacement, various operators
have to be commuted past each other and we have assumed that operators
built out of orthogonal linear combinations of $X$ and $Y$ commute. This
is true only for ``magic'' points of the magnetic field: as we discussed
in the preceding section, for other values of the field,
extra phases appear. Also, the above naive treatment of the breakup
of operators into left- and right-movers is strictly valid only for fields
compactified to the $SU(2)$ radius. The left and right moving momenta
will depend on the compactification radius, and, for this particular
radius only, left and right moving momenta are independent.
For any  radius, the Neumann state has zero total momentum ($ p_L = - p_R $)
and contains only winding modes. For infinite compactification radius there
are no winding modes (i.e. $p_L = p_R$) and the two conditions taken together
imply that the $R=\infty$ Neumann state has $p_L = p_R=0$. A key point is
that we can compute the infinite radius boundary state by first doing the
computation at the $SU(2)$ radius (as above) and then inserting  a projection
$P_\infty$ onto states states with $p_L = p_R$~\foot{At a finite multiple of
the SU(2) radius, $R = N \sqrt{2}$, the construction is similar but the
projection is onto states that have the allowed  winding number $p_L-p_R =
k N/ \sqrt{2}$.}:
\eqn\factortwo{
\ket{B_V} = \sec(\delta/2)~ P_\infty ~e^{i\betum {\cal R}_L}
        e^{-H_{pot}(2 X_L')-H_{pot}(2 Y_L')}\ket{N}_{SU(2)}~.}
In what follows, we will show how to make practical use of this formula.

Let us begin by studying the cylinder partition function between a Neumann
and an interacting boundary at infinite compactification radius. Using
the above transformation of the interacting boundary state, we can write
$$
Z^{NB_V} = \sec(\delta/2)
\bra N  q^{L_0 +\tilde{L}_0}  e^{i\betum  {\cal R}_L}
e^{-H_{pot}(2 X_L') -H_{pot}( 2 Y_L') }
\ket N
$$
The projector ${\rm P}_\infty$  onto $ p_L = p_R $ disappears because the
Neumann state $ \ket{N}$ has zero left and right moving  momenta.
Also, as we have no operators that change the right moving momentum,
only the zero momentum component of the state $\ket{N}_{SU(2)}$ contributes,
leaving just the usual Neumann boundary state $\ket N$.
We can perform the transformation
$X \ra X', Y \ra Y'$ everywhere in the previous equation,
since the Neumann states, the ${\cal R}_L$ operators and the
$L_0, \tilde{L}_0 $ operators are all invariant under rotations
in the $XY$-plane
(the rotation affects the left- and right-movers in the same way ).
The end result is a greatly simplified expression for the partition function:
\eqn\partfinal{
        Z^{NB_V} = \sec(\delta/2)
 \bra N  q^{L_0 +\tilde{L}_0} e^{i\betum  {\cal R}_L}
                e^{-H_{pot}(2 X_L) - H_{pot} ( 2 Y_L)} \ket N \ .}
The effect of the potential is to replace the Neumann boundary by a product
of two interacting $c=1$ boundary states which are then
mixed together in the partition function by the chiral $O(2)$ rotation
$e^{i\betum  {\cal R}_L}$. This simplification was possible because the $X$
and $Y$ potential operators effectively commute for the ``magic'' field
values satisfying $\tan(\betum /2)=n$, with $n\in\IZ$.

The fact that \partfinal\ is almost completely factorized into two exact
$c=1$ boundary states enables us to obtain an explicit expansion of the
partition function in powers of $q$. We exploit the fact that
the $c=1$ boundary
states are built on SU(2)$_1$ Kac-Moody algebras. We actually have two
commuting algebras, one for each boson, with the currents
\eqn\currents{\eqalign{
J^{ X \pm}(z) =  e^{\pm i 2 X(z) }\ ,~~~~~~~~~~~~~~~~
&J^{ Y \pm}(z) =  e^{\pm i 2 Y(z) }\ ,\cr
J^{X 3}(z) = { i  } \partial_{z} X(z)\ ,~~~~~~~~~~~~~~~~~~
&J^{Y 3}(z)=  { i  } \partial_{z} Y(z)\ .
}}
Using the results of \CKLM\ on the expression of the $c=1$ boundary state in
terms of Kac-Moody generators, we can rewrite the partition function
entirely in  terms of such generators:
\eqn\partcurrents{\eqalign{ Z^{NB_V}=&{\sec(\delta/2) }
\bra{0} \exp\bigl [ \sum_{n=1}^\infty {2 \over n} \{ \bar{J}_{n}^{X 3}
 ( \cos\betum J^{X 3}_{n}- \sin\betum J^{Y 3}_{n} ) +\cr
        &\bar{J}_{n}^{Y 3} ( \cos\betum J^{Y 3}_{n}
        + \sin\betum J^{X 3}_{n} ) \}\bigr ]
q^{L_0 +\bar{L}_0} \exp\bigl [ \sum_{n=1}^\infty {2\over n}
( \bar{J}_{-n}^{X 3} J_{-n}^{X \theta}+\bar{J}_{-n}^{Y 3} J_{-n}^{Y \theta} )
\bigr ]\ket{0}
} }
where the rotated current
$$
J^\theta_n = \cos\theta J^3_{n} + {\sin\theta \over 2 i}
( J^+_n - J^-_n )~,
$$
with $\theta=2\pi V$, contains all the information about the
strength of the boundary potential.
The commutation relations for the SU(2)$_1$ Kac-Moody algebra suffice to
compute the expansion of the partition function in powers of $q$. We obtain
\eqn\partresult{\eqalign{
Z^{NB_V}  = {1 \over \cos \delta/2 }  \{ 1 & + (2 \cos\theta
 \cos\betum) q^2 + \cr &
( 2 \cos^2\betum +\cos^2 \theta
\cos (2 \betum)  + 2 \cos\theta \cos \betum ) (q^2)^2 + \cdots~\}. }}
Each term in this expansion is exact in the parameters $ V $ and $\betum $
and requires only some $SU(2)_1$ algebra for its calculation. To reiterate,
this expression is only valid for the discrete field values where
$\tan(\betum /2)$ is an integer.

For the first non-trivial ``magic'' point, $(\alpha,\beta)=(1/2, 1/2)$,
it is possible to obtain a more explicit expression for the
partition function. This point is interesting because
it corresponds to a flux of $\pi$ through the unit cell of the
potential. The rotation by  $\delta=\pi/2$ amounts to the transformation
$X_L \ra Y_L, Y_L \ra - X_L $. We can commute $e^{i\betum  {\cal R}_L}$ past
$L_0$ so that it acts on the Neumann boundary. We now make use of the
expression of the Neumann boundary state as a sum over the discrete state
primaries $\ket{j,j_z}$ (explained in detail in \CKLM) to evaluate this action:
\eqn\rotket{\eqalign{ \bra N e^{i{\pi\over 2} {\cal R}_L} &=
\sum_{j,l} {\bra{ j,0} }^X_L{\bra{ j,0} }^X_R{\bra{ l,0} }^Y_L{\bra{ l,0} }^Y_R
 \sum_{N,M} \{N\}^X_L \{N\}^X_R  \{M\}^Y_L \{M\}^Y_R
        e^{i{\pi\over 2} {\cal R}_L} \cr
        &=\sum_{j,l} {\bra{ j,0} }^Y_L{\bra{ j,0 }}^X_R
        {\bra{ l,0} }^X_L{\bra{ l,0} }^Y_R
        (-1)^l \sum_{N,M} \{N\}^Y_L \{N\}^X_R  \{M\}^X_L \{M\}^Y_R }}
where  $\{N\}$ denotes a combination of Virasoro generators that
forms a basis of descendants for a Virasoro module. We note that the operator
${\cal R}_L$ does not commute with the Virasoro operators
of the fields $X$ and $Y$ (it commutes with the sum, of course), but
for this particular value of $\betum  $ we get $L^X \ra L^Y, L^Y \ra L^X $.

The product of $c=1$ boundary states can be similarly expanded using \CKLM:
\eqn\oneket{\eqalign{
{\ket B}^X {\ket B}^Y = \sqrt{2}
 \sum_{N',M'} \{N'\}^X_L&\{N'\}^X_R  \{M'\}^Y_L \{M'\}^Y_R \times\cr
&\sum_{j',l'} {\cal D}_{00}^{j'}(\theta) {\cal D}_{00}^{l'}(\theta)
{\ket {j',0}}^X_L {\ket {j',0}}^X_R{\ket{ l',0}}^Y_L{\ket {l',0}}^Y_R~.}}
The partition function is the sandwich of $q^{L_0 +\tilde{L}_0}$ between
\rotket\ and \oneket\ and can be expanded as
\eqn\bigmess{\eqalign{
Z^{NB_V}  = \sqrt{2}
\sum_{j,l,j',l'} \delta_{j,j'} \delta_{l,l'}\delta_{j,l'}\delta_{l,j'}&
\sum_{M,N,M',N'} \delta_{M,M'}  \delta_{N,M'} \delta_{M,N'} \delta_{N,N'}\cr
& {\cal D}_{00}^{j'}(\theta) {\cal D}_{00}^{l'}(\theta) (-1)^l
(q^2)^{\epsilon(j,N) + \epsilon(l,M)} }}
where $\epsilon(j,N)$ is the weight of the state $  \{N\} { \ket {j,0}}$.
Summing over all the delta functions and remembering the definition
$$
\sum_M (q^4)^{\epsilon(j,M)} = \chi^{Vir}_j(q^4) =
{q^{4 j^2} - q^{4 (j+1)^2 } \over q^{1\over 6} \prod_1^\infty (1 - q^{4 n} ) }
$$
of the Virasoro character, we obtain our final expression
for the partition function:
\eqn\infradius{
Z^{NB_V}  = \sqrt{2} \sum_{j=0}^{\infty} (-1)^j ( {\cal D}_{00}^{j}(\theta)
 )^2 \chi^{Vir}_j(q^4)~.
}
Once again, the boundary potential strength $V$ appears only in the $SU(2)$
rotation angle $\theta=2\pi V$.

\infradius\ is the answer for non-compact $X$ and $Y$, and it can easily be
extended to the compact case. If we choose the compactification radius
so that the fields $X', Y'$ (that appear after rotating the original fields
$X,Y$) are compactified at the $SU(2)$ radius, the compactified partition
function can be computed by the same methods as above with the result
\eqn\sutopfn{\eqalign{&
Z = \sqrt{2} \sum_{j=0,1/2,...}
\sum_m {\cal D}^j_{-m,m}(4 \pi V) e^{- i \pi j}
\chi^{Vir}_j(q^4) \cr
= & \sqrt{2} \sum_{j = 0,1/2,...} \chi^{SU(2)}(\hat\theta) \chi^{Vir}_j(q^4) =
{ 1 \over q^{1\over 6} f(q^4) }  \Theta_3({\hat\theta \over 4 \pi} | \tau )
}}
where $\hat\theta = 4 \pi V +\pi $,  $ q=e^{i\pi \tau} = e^{-2 \pi l/T}
$, $f(x) = \prod_1^\infty (1-x^n) $ and $\Theta_3$ is a Jacobi theta function.
 Apart from the replacement $q^2\to q^4$,
this is the same as the $c=1$ partition function at the $SU(2)$ radius.

As an interesting application of our results, we may study the effect
of magnetic field on the energy levels of DQM. For zero magnetic
field the partition function is a square of the $c=1$ partition function
which was evaluated in \CKLM\ for one Neumann and one dynamical
boundary,
$$ Z_{c=1}=
{1\over \sqrt{2}} \sum_{j=0}^{\infty} {\cal D}_{00}^{j}(2\pi V)
 \chi^{Vir}_j(q^2)~.
$$
Transforming to the open string channel, we find
$$ Z_{c=2}=Z^2_{c=1}={1 \over w^{1/12} f^2(w) }
\int_0^{2 \pi} {d\varphi_1 \over 2 \pi } \int_0^{2 \pi}
{d\varphi_2 \over 2 \pi }
\sum_{k = -\infty}^\infty \sum_{r = -\infty}^\infty
w^{( {\gamma_1  \over 4 \pi } +
k )^2+( {\gamma_2  \over 4 \pi } +r )^2 }
$$
where $w=e^{-\pi T/l}$ and
$$\cos{\gamma_1\over 2} = \cos (\pi V) \cos {\varphi_1}  \ ,
\qquad \cos{\gamma_2\over 2} = \cos (\pi V)
\cos {\varphi_2}\ .
$$
It is interesting to compare this formula with the corresponding
partition function at the first ``magic'' value of magnetic field.
Transforming \sutopfn\ to the open string channel, we find
$$\eqalign{
Z^{NB_V}  =& {1 \over w^{1/48} f(w^{1/2}) } {1\over \sqrt{2}}
\int_0^{2 \pi} {d\varphi_1 \over 2 \pi } \int_0^{2 \pi}
{d\varphi_2 \over 2 \pi }
\sum_{k = -\infty}^\infty w^{{1\over 2} ( {\gamma  \over 4 \pi } +
k )^2 }
\cr
=& {1 \over w^{1/12} f^2(w) }{1\over \sqrt{2}}
\int_0^{2 \pi} {d\varphi_1 \over 2 \pi } \int_0^{2 \pi}
{d\varphi_2 \over 2 \pi }
\sum_{k = -\infty}^\infty \sum_{r = -\infty}^\infty
w^{{1\over 2}( {\gamma  \over 4 \pi } +k )^2+
( {1\over 4 } +r )^2 }
} $$
where
$$
\cos {\gamma \over 2} = \sin(2\pi V) \sin{\varphi_1\over 2}
\sin{\varphi_2\over 2}~.
$$
This indicates that, near the tight-binding limit ($V=1/2$),
there are more bands per unit energy for $\delta=\pi/2$ than
for $\delta=0$. This is not surprising because $\delta=\pi/2$
corresponds to the non-trivial magnetic flux of $\pi$ per unit cell of the
square lattice.

\newsec{Exact Results From Fermionic Representation}

In the preceding section we obtained a simple representation of the
boundary state at the magic values of the magnetic field. We then
proceeded to calculate various partition functions and found that,
while exact results could be obtained, the
derivations were somewhat awkward.
For the $c=1$ theory, it was found that most results can also be obtained
quite simply by mapping the system into a theory of free fermions
\refs{\ghm,\PT}. It turns out that the same is true in the $c=2$ case.
In this section we review the $c=1$ fermionization and show how to extend
it to the $c=2$ system at the ``magic''  points. We will see that at all
these points the fermionic theory is essentially the same.

We begin by reviewing the mapping of the $c=1$ theory into a chiral free
fermion action \PT. We start with the lagrangian
\eqn\lag{ L={1\over 4\pi}\int_0^l d\sigma (\partial_\mu X)^2 - V \cos X(0)}
where the boundary at $\sigma=0$ is dynamical, while the boundary at $\sigma=l$
is free. We introduce chiral bosons in the usual way via
$ X(\sigma, t) = X_L (t+\sigma)+ X_R (t-\sigma)$. The Neumann boundary
conditions at $\sigma=0$ and $\sigma=l$
allow us to replace $X_L$ and $X_R$ by a
single left-moving field, periodic on the doubled interval:
$ X_L (\sigma+ 2l) = X_L (\sigma)$. The interaction term at $\sigma=0$ is
then replaced by $-V\cos (2 X_L)\vert_{\sigma=0}$.

In order to fermionize the theory,
Polchinski and Thorlacius introduced an auxiliary
boson $Y$. A pair of left-moving fermions were defined by
\eqn\fermions{   \psi_1 =C_1 e^{ i(Y_L- X_L)   }\ ,~~~~~~~~~~~~~~~~~~
\psi_2 =  C_2 e^{ i(Y_L+ X_L)  } }
where $C_i$ are the cocycles necessary to make the fermions
anticommuting.
The non-polynomial bosonic interaction term is mapped into
$$ L_{int} = -{V\over 2}
\left (\psi_1^\dagger \psi_2+
\psi_2^\dagger \psi_1\right )\vert_{\sigma=0}~.$$
There are some subtleties related to projection onto even fermion number,
and dividing by the partition function of the free auxiliary boson $Y$,
but this mapping into free fermions with quadratic interaction is a nice way
to bring out the hidden simplicity of the $c=1$ case.

Something very similar can be done for our $c=2$ system.
First consider the case of $\beta=0$ (no magnetic field).
The action \action\ reduces to two independent copies
of the $c=1$ theory \lag. The two chiral fermions may again be introduced
via \fermions, except that now $Y$ is a physical field, not an auxiliary
field. The $c=2$ interaction term
$-V_X\cos (2 X_L)-V_Y\cos (2 Y_L)\vert_{\sigma=0}$ is still a quadratic
in fermion language:
\eqn\nonc{ L_{int} = -{V_X\over 2}(\psi_1^\dagger \psi_2+
\psi_2^\dagger \psi_1)
-{V_Y\over 2} (\psi_1 \psi_2+ \psi_2^\dagger \psi_1^\dagger)
\vert_{\sigma=0}~.
}

It is interesting that when we turn on the magnetic field, exactly the same
fermionic theory specified by \nonc\ arises at all the ``magic'' points.
To show this, consider a string with opposite charges at the two ends
interacting with the constant magnetic field. The
magnetic interaction (written with Minkowski signature) becomes
$$ L_{mag} = {\beta\over 4\pi}\left[
        \left( X \partial_t Y
-Y \partial_t X \right)\vert_{\sigma=0}
        -\left( X \partial_t Y
-Y \partial_t X \right)\vert_{\sigma=l}
\right ]
\ .$$
The vanishing of boundary terms in the variation of the action leads to
identical boundary conditions at $\sigma=0$ and $\sigma=l$,
\eqn\sammag{ \partial_\sigma X-
{\beta\over \alpha} \partial_t Y=0\ ,
\qquad\qquad \partial_\sigma Y
+ {\beta\over \alpha} \partial_t X=0\ .
}
These boundary conditions induce the following relations between left-
and right-movers,
\eqn\reflect{\eqalign{&
 X_L - {\beta\over \alpha} Y_L = X_R + {\beta\over \alpha} Y_R\ ,\cr
 &Y_L + {\beta\over \alpha} X_L = Y_R - {\beta\over \alpha} X_R\ .
}}
These conditions allow us to show that $X_L$ and $Y_L$ are periodic with
period $2l$, and that the right-movers may be eliminated in favor of the
left-movers through
\eqn\elim{\eqalign{&
 X_R  = (\cos \delta) X_L - (\sin \delta) Y_L\ ,\cr
 &Y_R  = (\cos \delta) Y_L + (\sin \delta) X_L\ ,
}}
where $\delta$ is the rotation angle specified by \rotangle.
The interaction term then simplifies to
$$-V_X\cos (X_L+X_R) -V_Y\cos ( Y_L+Y_R))\vert_{\sigma=0}
= -V_X\cos ( 2X_L') -V_Y \cos ( 2Y_L'))\vert_{\sigma=0}
$$
where the primed fields are defined in \prf. It is easily checked that
the bulk action for $X'$ and $Y'$ at any ``magic'' point is normalized
identically to that of $X$ and $Y$ at the $\beta=0$ point. Hence,
subsequent fermionization gives the same result as the $\beta=0$
theory of \nonc.

It may come as a surprise that all the ``magic'' points are governed by the
same fermionic theory and, therefore, have the same cylinder partition
function. This is due to our particularly simple choice of boundary
conditions at $\sigma=l$ (recall that we endowed that end of the string
with a charge opposite to the one at $\sigma=0$).
This simplicity is also evident in the boundary state calculation of
the cylinder amplitude,
$$  Z^{B_0 B_V}  = \sec^2(\delta/2)
 \bra N  e^{-i\delta  {\cal R}_L} q^{L_0
 +\tilde{L}_0}e^{i\delta  {\cal R}_L}
 e^{-H_{pot} (2 X_L) -H_{pot}(2 Y_L) }
 \ket N
$$
Since ${\cal R}_L$ commutes with $L_0$, this reduces to
\eqn\decouple{
Z^{B_0 B_V}  =  \sec^2(\delta/2)
\left (\bra N_X q^{L_0 +\tilde{L}_0 }
 e^{-H_{pot} ( 2 X_L)} \ket N_X\right  )^2
}
which is simply the square of the $c=1$ partition, except for
the  factor  coming from the magnetic field contribution to
the boundary entropy. Thus, at any ``magic'' point we find the same
result as for the completely decoupled $\beta=0$ theory.

As we showed in the previous section,
matters are more complicated if
the boundary condition at $\sigma=l$ is simply Neumann. Then the
magnetic rotation does not cancel, and the
cylinder amplitude  depends on the ``magic'' point. It can be evaluated
explicitly only in the first non-trivial case ($\beta=\alpha$).
Here we repeat this calculation using the free fermion theory.

Since \elim\ still applies at $\sigma=0$, we may repeat the previous
transformations to arrive at the fermion coupling of \nonc.
Now separate the Dirac fermions into real and imaginary parts,
$\psi_1= \chi_{11}+ i\chi_{12}$, $\psi_2=\chi_{21}+i\chi_{22}$ so that
there are four left-moving Majorana fermions
$\chi^a=(\chi_{11},\chi_{12},\chi_{21},\chi_{22})$ and a corresponding
set of right-movers. The boundary interaction is a localized Majorana
mass term which in general has the effect of rotating the left-movers into
right-movers via an $SO(4)$ transformation
$\chi^a\rightarrow R^a_{I b} \chi_b$. It is not too hard to show that
the boundary interaction \nonc\ generates the particular $SO(4)$ rotation
$$ R_I=\pmatrix{ \cos\pi(V_X-V_Y) & 0 &- \sin\pi(V_X-V_Y)& 0 \cr
0& \cos\pi(V_X+V_Y) &0& -\sin \pi (V_X+V_Y)         \cr
\sin\pi(V_X-V_Y) & 0& \cos \pi (V_X-V_Y) &0       \cr
0&\sin\pi(V_X+V_Y) & 0& \cos\pi(V_X+V_Y) \cr} $$
We should also include the subsequent magnetic rotation of the boundary
state, $e^{i\delta  {\cal R}_L}$, which now
does not cancel.\foot{In open string language this
rotation introduces twisted boundary conditions in the $\sigma$ direction.}
For $\delta=\pi/2$, the magnetic rotation transforms the bosonic fields
according to
$$ X'_L \rightarrow Y'_L\ ,\qquad\qquad
Y'_L\rightarrow - X'_L\ .$$
In fermionic language this translates into
$\chi^a\rightarrow R^a_{M b} \chi_b$, where
$$ R_M=
\pmatrix{0&0&1&0 \cr 0&0&0&-1 \cr 1&0&0&0 \cr 0&1&0&0 \cr}$$
The overall rotation of the fermionic boundary state is $R= R_M R_I$.
It should be noted that the magnetic field acts locally on the fermions
only at $\delta=\pi/2$: at other magic values of the field, the
magnetic field rotation will not be a simple $SO(4)$ rotation.

In order to proceed we have to choose the boson compactification radius.
The simplest case is if we are at the SU(2) radius. From the
expressions for the fermions, \fermions\ we see that
the states created by them
have momenta $p^L_X=n$, $ p^L_Y=m$ with $n+m$ even.
In principle we could also consider states with $n+m$ odd. They
do not contribute, however; the magnetic field term  projects them out because
it interchanges $n \leftrightarrow m$ and the potential terms change $n$
and $m$ by two. Using the fermions,
the partition function at the SU(2) radius may be written as
$$
Z^{N B_V}_{SU(2)}= \sqrt{2}  (q^2)^{-2/24} \prod_{n=1}^{\infty}
\det( I+ q^{2n-1} R ) = $$
$$ \sqrt{2}  (q^2)^{-2/24} \prod_{n=1}^{\infty} (1 - q^{4n-2} )
(1 - 2 \sin\pi(V_X+V_Y)
q^{2n-1} + q^{4n-2} )
$$
which is equal to our previous result \sutopfn\
obtained using the bosonic theory
( with the replacement $ V \to (V_X + V_Y)/2 $ ).

To calculate the partition function for the infinite radius case
we should restrict to states with $n=m=0$, i.e.  zero fermion number for
both $\psi_1$ and $\psi_2 $. We can project onto states having zero
fermion numbers by inserting into the partition function a phase
$e^{i\varphi_1 F_1 +i\varphi_2 F_2 } $
and then integrating over $\varphi_i$.
The insertion of these phases corresponds to an additional SO(4) rotation
$$ R(\varphi) = \pmatrix{ \cos\varphi_1& \sin\varphi_1&0&0\cr
-\sin\varphi_1& \cos\varphi_1& 0&0 \cr
0&0& \cos\varphi_2 & \sin\varphi_2 \cr
0&0&-\sin\varphi_2 & \cos\varphi_2 \cr } ~.$$
The overall rotation is given by the matrix
$\tilde{R} \equiv  R(\varphi) R $,
and the partition function becomes
$$
Z^{N B_V}_{R=\infty}= \sqrt{2} \int_0^{2\pi}
{ d\varphi_1\over 2\pi} {d\varphi_2 \over 2 \pi}
 (q^2)^{-2/24} \prod_{n=1}^{\infty} \det( I+ q^{2n-1} \tilde{R} )
=$$
$$ \sqrt{2} \int_0^{2\pi} { d\varphi_1\over 2\pi} {d\varphi_2 \over 2 \pi}
 (q^2)^{-2/24} \prod_{n=1}^{\infty} (1-q^{4n-2} )(1 +
q^{2n-1}(\cos \varphi_1-\cos \varphi_2  ) \sin 2 \pi V
+ q^{4n-2}   )
$$
where we have, for simplicity, set $V_X=V_Y=V$. We have again
verified that this is identical to \infradius,
the result obtained using the bosonic theory.

\newsec{Boundary S-Matrix for Special Magnetic Field Values}

If the boundary is interpreted as an impurity in a bulk system, it is
often important to know the S-matrix for scattering from the impurity. As
is by now well-understood \refs{\afflud,\CKLM}, the S-matrix can be constructed
directly from the boundary state.  The boundary state encodes the information
on how the left movers are reflected as right movers. We can extract
the S matrix from the correlation function of left moving  $\partial X$ and
right moving  $\bar{\partial} X$ fields.
This reflection condition implies that we can express the right movers
in terms of left movers in the following way
\foot{This is the naive result of first carrying out the $SO(2)_L$ rotation
due to the magnetic field and then the independent $SU(2)_L$ rotations
due to the $c=1$ boundary states. It is correct as long as we are allowed
to ignore the commutation phases of the various dimension one operators.
As usual, this is true only for the ``magic'' values of magnetic field.}
\eqn\reflection{
\eqalign{
\bar{\partial} X \ra
\cos \theta   \p X' & - \cos \theta \tan(\delta/2)  \p Y' +\cr
& \sin \theta \big[  J^1 \big (X'\big) - \tan(\delta/2) J^1
\big(Y'\big)\big ] ,\cr
\bar{\partial} Y \ra
\cos \theta   \p Y' & + \cos \theta \tan(\delta/2) \p X' +\cr
&\sin \theta \big [  J^1 \big(Y'\big) + \tan(\delta/2)
J^1 \big(X'\big)\big ],
} }
where both sides are evaluated at $\bar{z}$, we use the  fields
\prf ~
and we define
$$J^1\big(X'\big)
= { e^{-2 i X' } -  e^{2 i X' } \over 2 }~,\qquad\theta = 2 \pi V~.$$
These formulae enable us to calculate any $S$-matrix
element. While the operators $\partial X$ describe localized excitations,
we find that, when they scatter off the boundary, solitons, described by
the exponentials in \reflection, are created. The discussion of
soliton sectors is parallel to that in \CKLM .

In applications to DQM, we are interested in calculating correlation functions
of the fields  $\dot X(t)$ and $\dot Y(t)$ at the boundary. These correlation
functions can be calculated from the bulk ones by taking the limit
\eqn\xdot{
\dot{X}(t) = \lim_{z,\bar{z} \rightarrow t}  (\partial X +
\bar{\partial} X ) ~.}
We can calculate in this way all two point correlation functions:
$$
\langle \dot{X}(t_1) \dot{X}(t_2) \rangle =
  { - 2 \big [ \cos^2(\theta/2) + \sin^2(\theta/2) \tan^2(\delta/2)
\big ] }
 {1 \over (t_1-t_2)^2},
$$
the same result for $ \langle \dot{Y} \dot{Y} \rangle $, and
$$
\langle \dot{X}(t_1) \dot{Y}(t_2) \rangle = -2 i  \cos \theta \tan(\delta/2)
 \pi \delta'(t_1 - t_2)~
$$
which is a contact term.
These correlation functions satisfy the duality relations
postulated in \cgcdef, relating  their behavior at
the different ``magic'' points.

Now we consider higher point functions.
As usual we will be interested
in the connected parts. If we focus our attention on
the long time behavior we can simplify the calculations by
 taking  the limit in \xdot  ~  before we calculate
the correlators. Using \reflection ~   we get
\eqn\transxt{
\dot{X} = 2  \big [\p X' + J^\theta\big(X'\big) \big ] +
2 \tan(\delta/2)\big [\p Y' - J^\theta\big(Y'\big) \big],
}
\eqn\transyt{
\dot{Y} = 2 \big [\p Y'+ J^\theta\big(Y'\big)\big ] -
2 \tan(\delta/2)\big [\p X' - J^\theta\big(X'\big)\big ]
}
where we have defined
$$
J^\theta\big(X'\big) \equiv
\cos \theta \p X' + \sin \theta J^1\big(X'\big)\ .
$$
Now we will show that correlation functions of only $\dot{X}$s have no
connected piece. To that effect we perform a sequence of
 field redefinitions to get
$$
\dot{X} = N \p U\ ,
$$
$$
\dot{Y}=- N[ (\cos\gamma) J^1(W\cos\gamma  + U\sin\gamma  )
+ (\sin\gamma) J^1( U\cos\gamma  - W\sin\gamma  ) ]\ ,
$$
where $U, W$ are uncorrelated free fields with the same normalization
as the  fields $X',~Y'$ and we have defined
$$
 N= 2\left[ \cos^2(\theta/2) + \sin^2(\theta/2) \tan^2(\delta/2)
\right]^{1/2}\ ,\qquad\qquad
\tan \gamma = \tan (\delta/2) \tan (\theta/2)  ~.
$$

For the case in which we only have $\dot{X}$s the higher-point correlation
functions reduce to the free correlation functions except for an overall
factor.
$$
\langle \dot{X}(t_1) \cdots \dot{X}(t_{2n}) \rangle =
N^{2n} \langle \p U(t_1) \cdots \p U(t_{2n}) \rangle
$$
Thus correlation functions of only $\dot{X}$s  have no connected parts
despite the interactions. This is precisely the situation we had for the
single field case \CKLM .  The same is, of course, true for correlation
functions
involving only $\dot{Y}$s. We have also calculated mixed correlation functions
with two $\dot Y$s, finding the following connected piece
$$\eqalign{
\langle \dot{X}(t_1)  \cdots \dot{X}(t_{2n})& \dot{Y} (s_1)\dot{Y} (s_2)
\rangle_{connected} = {(-1)^{n+1} \over 2}   N^{2n+2}
        \cos^2 \gamma \sin^2 \gamma \times\cr
& [ (\sin\gamma)^{2n-2} + (\cos \gamma )^{2n-2} ] {(s_1-s_2)^{2n-2}
\over \prod_1^{2n} (t_i-s_1)  (t_i-s_2) } }
$$
These quantities have previously been calculated perturbatively by Freed
\freed\ and our results agree with hers up to a factor of 2.
We could go further and calculate more general connected correlators, but
little new would be learned. If the potential is turned off ($V\to 0$) or made
infinitely strong $(V \to {1\over 2}$), all the connected higher point
functions
vanish, as expected. What is remarkable is that for intermediate values of
the potential strength, the connected n-point functions are, apart from
overall constants, simple rational functions of coordinate differences.
This is in accord with general arguments by Freed in the work cited above.
The momentum-space S-matrix elements are correspondingly piecewise linear
functions of the scattering particle energies. Presumably this would not be
true at the perturbative fixed points that govern the behavior away from the
``magic'' magnetic field values. It would be instructive to compute scattering
data in those theories.

\newsec{Conclusions}

In this paper we have studied a class of $c=2$ boundary field theories whose
partition function reduces to a one-dimensional Coulomb gas with charges
obeying fractional interchange statistics. The typical context for this theory
is the dissipative quantum mechanics of a particle moving in a two-dimensional
periodic potential  and a magnetic field. The size of the magnetic field
determines
the statistics phase. This phase has a very significant effect on the
renormalization group flow of the theory and we have devoted this
paper to identifying and exploring the fixed points of these flows.

For certain discrete values of the magnetic field, the statistics phase is
unity and it turns out that we can find the exact boundary state describing
the associated conformal field theory. It is related in a simple
way to the exact boundary state which governs the $c=1$ version of this
problem (dissipative quantum mechanics of a particle moving in a periodic
one-dimensional potential with no magnetic field). Thus, the boundary state
and all other physical quantities are built out of $SU(2)$ group rotations
(even though strict $SU(2)$ symmetry is broken by the potential itself).
We have derived explicit formulae for partition functions and S-matrix
elements at the lowest non-trivial special value of the magnetic field.

When the statistics phase is not unity, we find a renormalization group
flow to a non-trivial fixed point that is reliable (perturbative)
for suitably chosen $\alpha$ and $\beta$. In these cases we
find that there is a nonzero boundary entropy and we can verify that it
decreases along the renormalization group flow. It should be feasible
and instructive to use open string field theory techniques to compute
the resulting zero-temperature entropy as a string field theory action.

The most promising potential application of the type of boundary field
theory we have discussed here is edge current tunneling in the quantum
Hall effect. The $c=1$ boundary field theory, in particular, has found
spectacular application \fls\ in the description of edge tunneling
in the $\nu=1/3$ fractional quantum Hall effect. These applications
rely in an essential way on the remarkable fact that the
$c=1$ boundary sine-Gordon model flows toward the conformal fixed point through
a sequence of {\it integrable} theories \gz. It is tempting to speculate
that the type of theory discussed in this paper will have similar application
to the quantum Hall problem at some other filling factor or more complex
geometry. A necessary precondition for this is presumably to find
integrable models which flow to our conformal theories. We have not seen
how to do this, but we hope that this work will stimulate others to try.

\vskip .3in
\centerline{\bf Acknowledgements}
\vskip .3in

This work was supported in part by DOE grant DE-FG02-91ER40671.
The work of I. R. K. was also supported in part by NSF Presidential Young
Investigator Grant No. PHY-9157482, James S. McDonnell Foundation Grant
No. 91-48 and the A. P. Sloan Foundation.

\listrefs

\bye